\documentclass[
 reprint,
jcp,
floatfix,
]{revtex4-1}

\usepackage{graphicx}
\usepackage{color}
\usepackage{enumitem}
\usepackage{amsmath}
\usepackage{multirow}
\usepackage{booktabs}
\usepackage{amsbsy}

\begin{document}

\def\homo{\mbox{\scriptsize HOMO}}

\def\sla{$^1L_a$}
\def\slb{$^1L_b$}
\def\tla{$^3L_a$}
\def\tlb{$^3L_b$}

\graphicspath{{figures/}}

\title{An assessment of low-lying excitation energies and triplet instabilities of organic molecules with an ab initio Bethe-Salpeter equation approach and the Tamm-Dancoff approximation}

\author{Tonatiuh Rangel}
\affiliation{Molecular Foundry, Lawrence Berkeley National Laboratory, Berkeley, California 94720, United States}
\affiliation{Department of Physics, University of California, Berkeley, California 94720, United States}
\email{trangel@lbl.gov}

\author{Samia M. Hamed}
\affiliation{Molecular Foundry, Lawrence Berkeley National Laboratory, Berkeley, California 94720, United States}
\affiliation{Department of Physics, University of California, Berkeley, California 94720, United States}
\affiliation{Department of Chemistry, University of California, Berkeley, California 94720, United States}

\author{Fabien Bruneval}
\affiliation{Service de Recherches de M\'etallurgie Physique, CEA, DEN, Universit\'e Paris-Saclay, F-91191 Gif-sur-Yvette, France}
\affiliation{Molecular Foundry, Lawrence Berkeley National Laboratory, Berkeley, California 94720, United States}
\affiliation{Department of Physics, University of California, Berkeley, California 94720, United States}

\author{Jeffrey B. Neaton}
\affiliation{Molecular Foundry, Lawrence Berkeley National Laboratory, Berkeley, California 94720, United States}
\affiliation{Department of Physics, University of California, Berkeley, California 94720, United States}
\affiliation{Kavli Energy Nanosciences Institute at Berkeley, Berkeley, California 94720, United States}
\email{jbneaton@lbl.gov}
\begin{abstract}
The accurate prediction of singlet and triplet excitation energies is of significant fundamental interest and critical for many applications. 
An area of intense research, most calculations of singlet and triplet energies use time-dependent density functional theory~(TDDFT) in conjunction with an approximate exchange-correlation functional.
In this work, we examine and critically assess an alternative method for predicting low-lying neutral excitations with similar computational cost, the ab initio Bethe-Salpeter equation (BSE) approach, and compare results against high-accuracy wavefunction-based methods. 
We consider singlet and triplet excitations of 27 prototypical organic molecules, including members of Thiel's set, the acene series, and several aromatic hydrocarbons exhibiting charge-transfer-like excitations. 
Analogous to its impact in TDDFT, we find that the Tamm-Dancoff approximation (TDA) overcomes triplet instabilities in the BSE approach, improving both triplet and singlet energetics relative to  higher level theories. 
Finally, we find that BSE-TDA calculations built on effective DFT starting points, such as those utilizing optimally-tuned range-separated hybrid functionals, can yield accurate singlet and triplet excitation energies for gas-phase organic molecules. 
\end{abstract}

\maketitle
\section{Introduction}

The quantitative prediction and understanding of low-lying excitations in organic molecules is of significant fundamental interest and technological relevance.
For example, a better understanding of multiexciton phenomena in organic molecular systems – such as singlet fission~(SF)~\cite{anthony_larger_2008,smith_recent_2013,lee_singlet_2013}, a process by which a singlet exciton decays into two low-energy triplet excitations, can lead to external quantum device efficiencies above 100\% \cite{smith_recent_2013,lee_singlet_2013} and is therefore desirable for next-generation solar cells and other optoelectronic applications. 
Such multiexciton energy conversion phenomena are dependent on a subtle balance between singlet and triplet excitation energies, and predictions of such energetics call for accurate ab initio methods.

A widely-used ab initio formalism for neutral excitations is time-dependent density-functional theory~(TDDFT). 
For gas-phase acene molecules, the performance of TDDFT with a number of exchange-correlation functionals is well-documented: overall, TDDFT with standard functionals---e.g., local, semilocal, and global hybrid exchange-correlation functionals---fails to predict triplet excitations~\cite{peach_influence_2011,peach_overcoming_2012} by $0.4$--$1.8$~eV, as well as the ordering and absolute energies of the two lowest-lying singlets~\cite{grimme_substantial_2003,kuritz_charge-transfer-like_2011}, one of which has charge-transfer-like character~\cite{kuritz_charge-transfer-like_2011}~(as detailed in Section~\ref{sect:la-lb}). \
These failures have been ascribed to i) the so-called ``low orbital overlap problem'' in global hybrid functionals, in which the overlap between spatially-separated molecular orbitals is usually overestimated; and to ii) triplet instabilities associated with TDDFT using standard approximate exchange-correlation functionals~\cite{kuritz_charge-transfer-like_2011,lopata_excited-state_2011,sears_communication:_2011,casida_progress_2012,peach_overcoming_2012}.

\begin{figure}[t]
\includegraphics[width=1.0\linewidth]{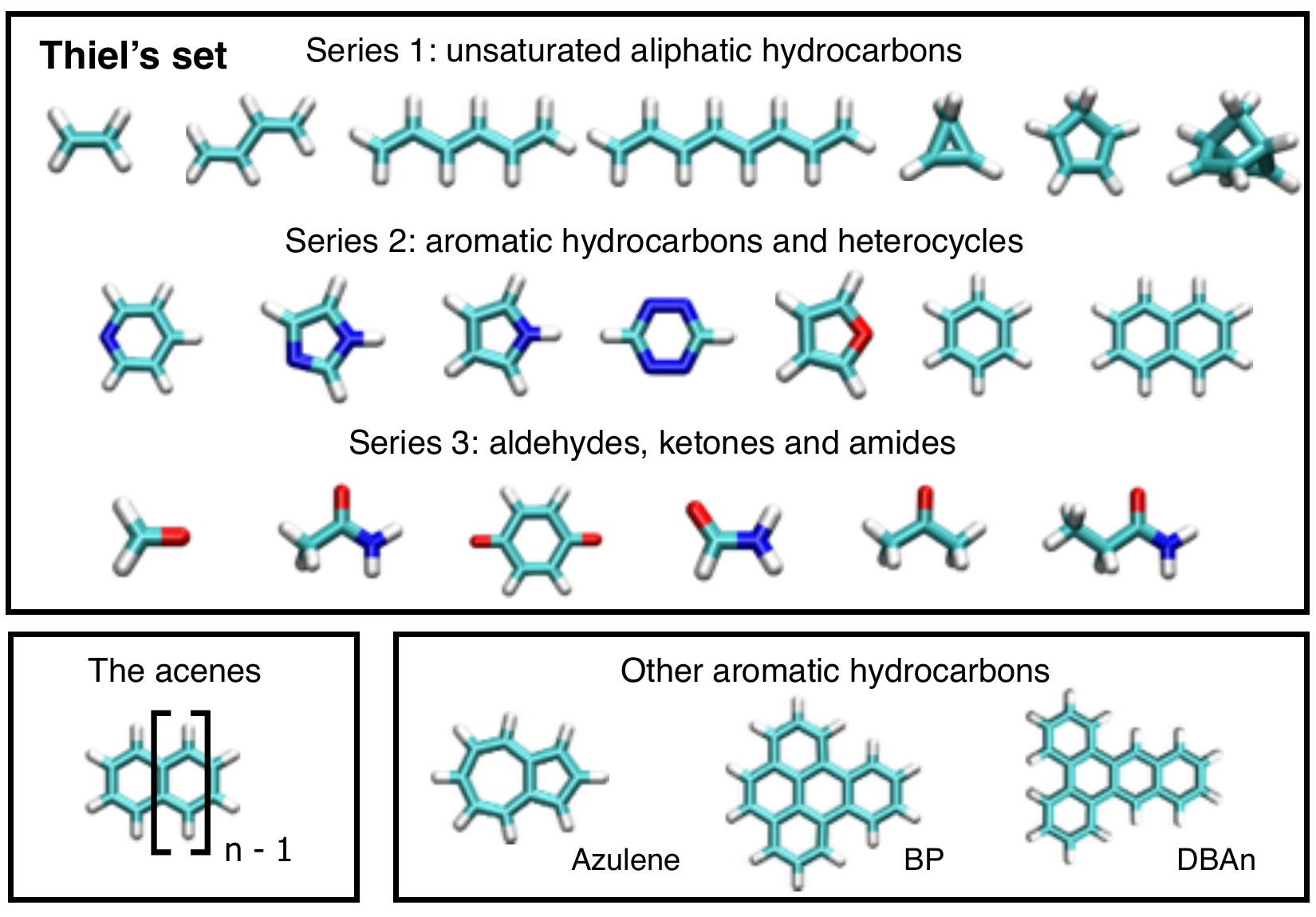}\\
\caption{
Top: Subset of 20 organic molecules containing triplet excitations from the Thiel's set.
Bottom: The general formula for an acene molecule, and the three other aromatic hydrocarbons stuided here: azulene,
benzo[$e$]pyrene~(BP) and dibenzo[$a,c$]anthracene~(DBAn). 
H is white, C is light blue, N is dark blue and O is red.
}
\label{fig:molecules}
\end{figure}

Beyond TDDFT approaches with conventional functionals, range-separated hybrid functionals~(RSH) have been shown to mitigate the low orbital overlap problem~\cite{wong_optoelectronic_2010,richard_time-dependent_2011,kuritz_charge-transfer-like_2011,peach_overcoming_2012}.
In this class of functionals, the Coulomb potential is partitioned into short- and long-range contributions, with the important consequence that different fractions of exact exchange can be used in the short and long range~\cite{leininger_combining_1997,yanai_new_2004}.
This partitioning is usually described as
\begin{equation}
\frac{1}{r} = \frac{\alpha + \beta \textrm{erf}(\gamma r)}{r} 
+  \frac{1-\left[\alpha + \beta \textrm{erf}(\gamma r)\right]}{r}, 
\end{equation}
where the first term is treated explicitly, and the second is replaced with a semi-local functional, such as one of several generalized gradient approximations~(GGAs).
The $\alpha$, $\beta$ and $\gamma$ parameters are either fixed as in, e.g., CAM-B3LYP~\cite{yanai_new_2004}, or tuned to fulfill DFT theorems as in optimally-tuned range-separated hybrid~(OTRSH) functionals~\cite{kronik_excitation_2012} or Koopmans’ compliant functionals~\cite{borghi_koopmans-compliant_2014}. 
Two examples of OTRSH functionals are that of Baer-Neuhauser-Lifshitz~(BNL)~\cite{baer_density_2005,livshits_well-tempered_2007} and the Perdew-Burke-Ernzerhof~(PBE)-based OTRSHs~\cite{kronik_excitation_2012,refaely-abramson_quasiparticle_2012,refaely-abramson_gap_2013}.
Importantly, CAM-B3LYP, OTRSH-BNL, and other RSH functionals have proven quite successful in predicting the low-lying excitations of aromatic hydrocarbons~\cite{wong_optoelectronic_2010,richard_time-dependent_2011,kuritz_charge-transfer-like_2011,lopata_excited-state_2011,peach_overcoming_2012} and charge-transfer~(CT) excitations~\cite{stein_prediction_2009,autschbach_charge-transfer_2009,stein_reliable_2009}.

An alternative approach to neutral excitations is ab initio many-body perturbation theory (MBPT), where neutral excitations are computed via two-particle Green's functions through solution of an effective two-particle equation, the Bethe-Salpeter equation~(BSE). 
In MBPT, solutions to the BSE build on one-particle energies and wavefunctions, usually obtained from a generalized Kohn-Sham DFT starting point within $GW$ approximation, where $G$ is the one-particle Green's function and $W$ the screened Coulomb interaction.
Referred to as the $GW$-BSE approach hereafter, this method has been extremely successful for solids~\cite{onida_electronic_2002}, as it goes beyond DFT by including electron-hole interactions, which can be significant for molecules and other systems. 
It has also been successfully applied to gas-phase molecules~\cite{onida_reining_1995,grossman_rohlfing_2001}, and quantitative and extensive benchmark studies are beginning to appear~\cite{boulanger_fast_2014,bruneval_systematic_2015,jacquemin_benchmarking_2015,jacquemin_00_2015,jacquemin_benchmark_2017}.
Yet much remains unknown about the performance of ab initio $GW$-BSE calculations, particularly their ability to predict acene excitations, charge-transfer-like excitations of aromatic molecules, and more generally, the triplet excitations of organic compounds. 
The aim of the present work is to address these issues.

The Bethe-Salpeter equation is a formal solution to the two-particle Green's function, giving access to excitonic wavefunctions and eigenvalues. 
The underlying theory and approach are explained in more detail in Refs.~\cite{onida_electronic_2002,rohlfing_electron-hole_1998,strinati_application_2008}. 
In finite systems with real wavefunctions, BSE is exactly analogous to TDDFT. 
As in the Casida equations of linear-response TDDFT, solutions to the BSE can take the form of an eigenvalue problem, i.e.,
\begin{equation}
\begin{pmatrix}
\phantom{-}A & \phantom{-}B\\
-B & -A
\end{pmatrix}
\begin{pmatrix}
X_s \\
Y_s
\end{pmatrix}
=
\Omega_s
\begin{pmatrix}
X_s \\
Y_s
\end{pmatrix},
\label{eqn:hamiltonian}
\end{equation}
where $X_s$, $Y_s$ are the excitonic wavefunctions, $\Omega_s$ are the eigenvalues, and the $A$ and $B$ blocks form the resonant and coupling parts of the BSE respectively. ~\cite{rohlfing_electron-hole_1998}.
In the Tamm-Dancoff approximation~(TDA)~\cite{_quantum_2003} the $B$ and $-B$ blocks are neglected, resulting in the decoupling of the $A$ and $-A$ blocks.
While the applicability and implications of the TDA in TDDFT are well documented~\cite{hirata_time-dependent_1999,casida_charge-transfer_2000,peach_influence_2011,peach_overcoming_2012,sears_communication:_2011,casida_progress_2012}, the quantitative impact of the TDA on $GW$-BSE calculations of small molecules remains underexplored.

Benchmarks against wavefunction-based theory are the norm to assess the accuracy of approximations within lower order theories or approximations to TDDFT.
For example, Thiel's set~\cite{silva-junior_benchmarks_2010} contains reference-quality excitation energies which were obtained with multistate multiconfigurational second-order perturbation theory such as (MS-CASPT2) and various coupled-cluster theories, such as coupled cluster with singles, doubles, and perturbative triples (CCSD(T))~\cite{silva-junior_benchmarks_2010,silva-junior_basis_2010}.
Recent work~\cite{bruneval_systematic_2015,jacquemin_benchmarking_2015,jacquemin_benchmark_2017} has explored the performance of $GW$-BSE for Thiel's set, reporting that $GW$-BSE can indeed yield quantitative singlet energetics under several approximations.
However, as in TDDFT, $GW$-BSE-calculated triplets were found to be substantially underestimated~\cite{bruneval_systematic_2015,jacquemin_benchmark_2017}.
The limited performance of $GW$-BSE in this case merits further analysis.

Motivated by the success of $GW$-BSE in general, and its low computation cost relative to wavefunction methods, herein we assess the performance of different approximations to $GW$-BSE and determine successful approaches within this framework for the quantitative prediction of low-lying excitations of organic compounds. 
We evaluate $GW$-BSE against multireference and coupled cluster references for representative singlet and triplet excitations of 27 organic molecules, including hydrocarbons, heterocycles, aldehydes, ketones, and amides (see Figure~\ref{fig:molecules}).
We focus on approximations to $GW$ that enter the BSE, including hybrid functional-starting points with one-shot schemes and the effect of partially self-consistent schemes, as described in Section~\ref{sect:comput-details}. 
We also provide a detailed assessment of the performance of the BSE and the TDA relative to that of other two-particle Green's function approaches and computationally-less-expensive TDDFT methods.

\section{Computational details}
\label{sect:comput-details}
Our calculations start with a self-consistent time-independent DFT calculation, using an approximate exchange-correlation functional (see below). 
For the molecules considered here, we minimize the total energy with respect to the density using fixed atomic coordinates for all molecules obtained from Ref.~\cite{silva-junior_benchmarks_2010} (see SI for more details). 
Starting from the output of our DFT calculations, and using the \textsc{MolGW} package~\cite{bruneval_molgw_2016,molgw_cite}, we then compute one- and two-particle excitation energies with the $GW$ and $GW$-BSE approaches, respectively.  
As is standard, our $GW$-BSE calculations build on single-particle states, which are coupled in the two-particle BSE equation via the electron-hole interaction kernel. 
With $GW$ input, BSE is recast into an eigenvalue problem~\cite{rohlfing_electron-hole_1998}, the solution of which yields the energies and eigenstates of a set of neutral excitations.
As detailed in prior work by us and others~\cite{x._blase_first-principles_2011,bruneval_benchmarking_2013,gallandi_accurate_2016,knight_accurate_2016,rangel_evaluating_2016}, $GW$ calculations are sensitive to the generalized Kohn-Sham starting point and to whether self-consistency is used.
Here, we build on previous work and use three accurate $GW$ schemes: 
$G_0W_0$@BHLYP (one-shot GW  on top of BHLYP~\cite{foresman_exploring_1996}, which has 50\%  exact-exchange); $G_0W_0$@OTRSH-PBE~\cite{perdew_generalized_1996}; and eigenvalue self-consistent $GW$~(ev$GW$), in which the quasiparticle energies are updated (in both $G$ and the polarizability) one or more times prior to calculating the final self-energy corrections~\cite{x._blase_first-principles_2011}.

As mentioned, our $GW$-BSE calculations are performed with the \textsc{MolGW} package, in which the frequency dependence of the $GW$ non-local self-energy $\Sigma(\mathbf{r},\mathbf{r'},\omega)$ is treated analytically, and hence is exact for a given basis set, without the need for plasmon-pole approximations.
We use conventional approximations to solve the BSE:
irreducible vertices are set to 1, 
the polarizability and other matrix elements are constructed using $GW$ eigenvalues and DFT wavefunctions, 
the screened Coulomb interaction is evaluated in the random phase approximation~(RPA), and a static electron-hole screening (COHSEX) is used; see Ref.~\cite{onida_electronic_2002}. 
We adopt the aug-cc-pVTZ basis set~\cite{dunning_jcp1989} which ensures convergence better than 0.1~eV for the excitation energies shown here (see SI for details).
In order to reduce the computational load, and for the purpose of parallelization, we use the resolution-of-the-identity in the Coulomb metric~\cite{vahtras_cpl1993,weigend_pccp2002}, as implemented in \textsc{MolGW}~\cite{rangel_evaluating_2016,bruneval_molgw_2016}, with the well-established auxiliary basis sets of Weigend~\cite{weigend_pccp2002} which are consistent with Dunning basis sets.
The resolution-of-the-identity is expected to have a small affect on the $GW$ energies, on the order of 1~meV, as we have demonstrated in the case of benzene~\cite{rangel_evaluating_2016}. 

For OTRSH-PBE, as a standard procedure for the acenes~\cite{refaely-abramson_quasiparticle_2012,refaely-abramson_gap_2013,rangel_evaluating_2016}, we set $\alpha = 0.0 - 0.2$, (see SI) which fixes the amount of short-range Fock exchange to $0-20$\%. 
Additionally, we set $\alpha + \beta = 1$ to enforce long-range asymptotic exact exchange.
Then, the range-separation parameter $\gamma$ is varied to achieve a minimization of the target function
\begin{eqnarray}
  J^2(\gamma) & = & \left[\mathrm{IP}^\gamma(N) + E^\gamma_{\homo}(N)\right]^2 \nonumber\\
 & + & \left[\mathrm{IP}^\gamma(N+1) + E^\gamma_{\homo}(N+1)\right]^2,
\end{eqnarray}
where the ionization potential of the neutral species with $N$ electrons, $\mathrm{IP}^{\gamma}(N)$, is determined via a $\Delta SCF$ approach from total energy differences as $\mathrm{IP}^{\gamma}(N)=\epsilon^{\gamma}_\textrm{tot}(N-1)-\epsilon^{\gamma}_\textrm{tot}(N)$.
Here $\epsilon^{\gamma}_{\textrm{tot}}(N)$ and $\epsilon^{\gamma}_{\textrm{tot}}(N-1)$ are total energies of the neutral and cation species respectively.
This procedure enforces the ionization potential theorem of DFT~\cite{perdew_density-functional_1982,salzner_koopmans_2009,levy_exact_1984,perdew_comment_1997,almbladh_exact_1985}, namely that the energy of the Kohn-Sham highest occupied molecular orbital~(HOMO) is equal to the negative of the first ionization potential.   
For molecules with unbound $N+1$ anionic state, only the first of these two terms is minimized, as in our previous work~\cite{rangel_evaluating_2016}.
The optimal parameters obtained within this framework for the molecules studied are listed in the SI. 
OTRSH-BNL parameters are taken from Ref.~\cite{kuritz_charge-transfer-like_2011}.

Our TDDFT calculations are performed with QChem~4.2~\cite{qchem} with standard settings, excluding core electrons in the correlation computation and neglecting relativistic effects as usual.
We use the cc-pVTZ basis set which, relative to aug-cc-pVTZ, converges the neutral-excitation energies satisfactorily: we consider TD-CAM-B3LYP, with and without the TDA, for the singlet $L_a$ and $L_b$ states for all acenes considered herein, and the lowest lying triplet state for benzene, naphthalene and anthracene. For all test cases, the difference between the augmented and unaugmented bases was between 0.001 and 0.087~eV, with an unsigned average difference of 0.028~eV.  

\section{Low-lying $\pi \rightarrow \pi^*$ excitations of aromatic hydrocarbons}
\label{sect:la-lb}
In aromatic hydrocarbons, like the acenes, azulene, benzo$[e]$pyrene~(BP) and dibenzo$[a,c]$- anthracene~(DBAn) (see Figure~\ref{fig:molecules}), the two low-lying singlet excitations are  labeled $^1L_a$ and $^1L_b$~\cite{platt_classification_1949}.
These excitations are well-known to differ significantly in character,  and these differences are widely discussed in the literature.
The bright (or large oscillator strength) longitudinal ionic \sla\ state involves principally a transition between the highest-occupied molecular orbital~(HOMO) and the lowest unoccupied molecular orbital~(LUMO), and is often described as having charge-transfer~(CT)-like character; 
while the dark (near-zero oscillator strength) covalent \slb\ excitation arises from a destructive interference~\cite{guidez_origin_2013} of transitions that typically couple the HOMO to the LUMO$+1$ and the HOMO$-1$ to the LUMO~\cite{hirao_complete_1997,parac_tddft_2003,kuritz_charge-transfer-like_2011}. 

The description of these excitations as ionic or covalent comes from valence bond theory, and refers to the distribution of charge in the spatial part of the component orbitals of the excited state. If in the resonance structures describing these orbitals, the density oscillates from negative to positive with respect to the carbon-atom centers, the excitation is termed ``ionic''. If there is no such oscillation and the resonance structures correspond to Kekule structures, with alternating double and single bonds, the excitation is termed ``covalent''. 


The corresponding low-lying triplet excitations are labeled, following the same conventions as above, as \tla\ and \tlb, respectively. 
Notice that labeling of unbound molecular orbitals (e.g. the LUMO and LUMO$+1$ of some acenes) is somewhat arbitrary, since their ordering may change depending on the choice of DFT exchange-correlation~(XC) functional and basis set. 
In this work, we adopt a definition of the unbound LUMO of benzene and naphthalene as the first resonant state whose energy corresponds to the negative electron affinity energy measured in experiments, as detailed in Ref.~\cite{hajgato_benchmark_2009}. 
As this state is not the most stable, the resonant state has a finite lifetime. 
In this work, we first computed the LUMO within PBE.
PBE spuriously binds the LUMO, however the corresponding wavefunction will only serve as a basis function to identify the LUMO within the more accurate approximations. We then define the LUMO calculated with a hybrid functional as the unbound state having the largest overlap with the PBE LUMO. 
With this simple method we have been able to extract the LUMO states across the acene series in a consistent and reliable manner.

Predicting both \sla\ and \slb\ presents a challenge for TDDFT approaches. 
In fact, \sla\ excitations, with their CT-like character~\cite{kuritz_charge-transfer-like_2011}, are usually poorly predicted by standard TDDFT~\cite{parac_tddft_2003,kuritz_charge-transfer-like_2011,moore_charge-transfer_2015},
due to the known shortcoming of many standard functionals to describe such excitations. We note that this shortcoming has the potential to be ameliorated by using RSHs with asymptotic exact exchange~\cite{kuritz_charge-transfer-like_2011},
although the CT nature of \sla\ excitations and the ability of RSHs to overcome these shortcomings has been questioned.~\cite{moore_charge-transfer_2015}.

\section{Results and discussion}
\label{results}

We begin with a benchmark of $GW$-BSE and TDDFT against CCSD(T) for the low-lying singlet excitations of the acenes.
We end with an examination of the role of the TDA within both the TDDFT and $GW$-BSE frameworks.
\subsection{Predicting the low-lying excitations of the acenes with TDDFT}
\label{tddft}

\begin{figure*}[t]
\includegraphics{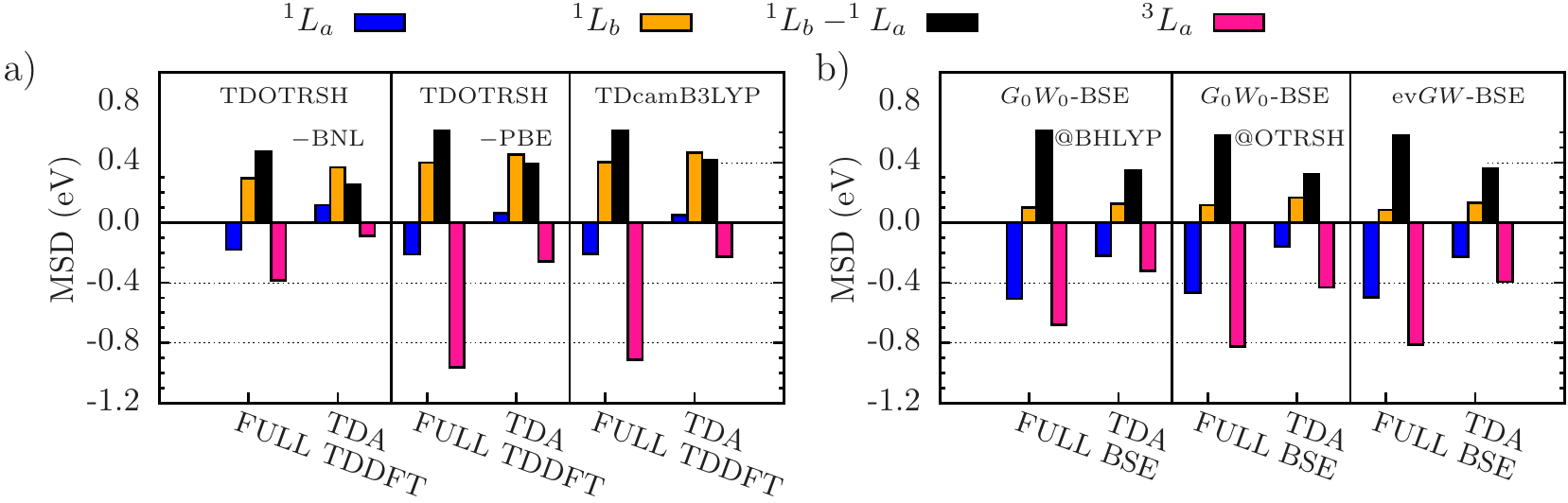}\\
\caption{MSD (see text for details) with respect to CCSD(T) ~\cite{hajgato_benchmark_2009,lopata_excited-state_2011} of calculated neutral excitations of the acene molecules ($n=1$ to $6$).
The calculated \sla\ (blue bars), \slb\ (orange bars), and \tla\ (pink bars) excitations, and \slb$ - $\sla\ (black bars) energy difference are shown for a few representative TDDFT and $GW$-BSE approaches:
TD-OTRSH and TD-CAM-B3LYP in panel a, and $G_0W_0$-BSE@BHLYP, $G_0W_0$-BSE@OTRSH-PBE and ev$GW$-BSE@PBE0 in panel b. 
}
\label{fig:mae-exct-tda}
\end{figure*}

\begin{figure*}[t]
\begin{tabular}{cc}   
\includegraphics{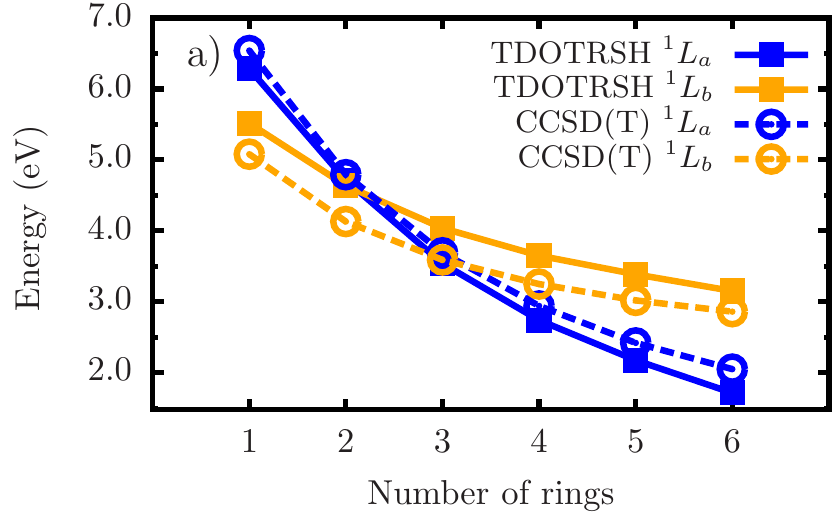}
\includegraphics{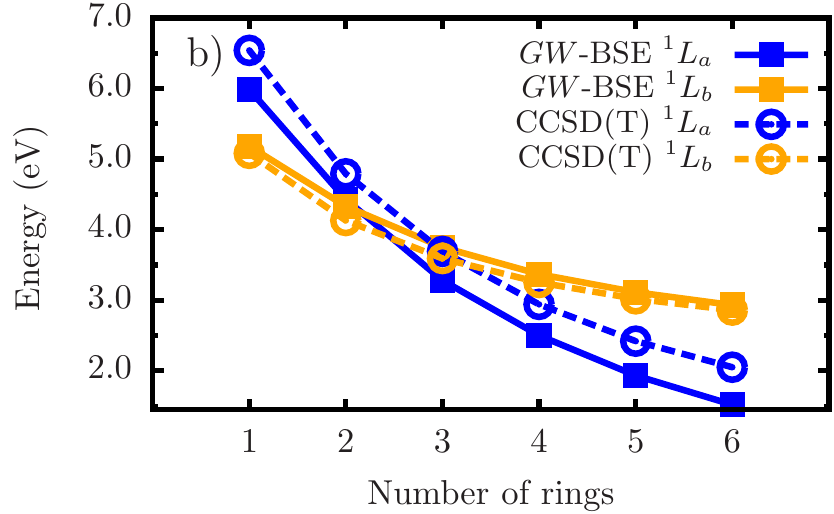}
\\
\includegraphics{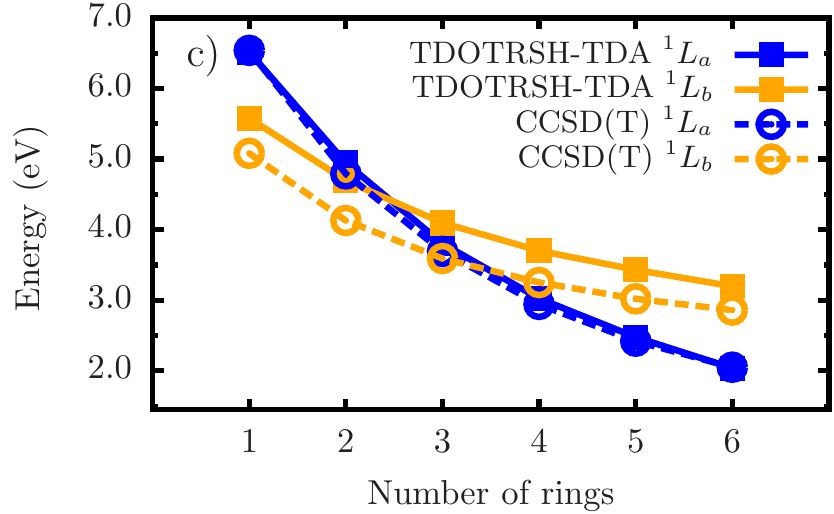}
\includegraphics{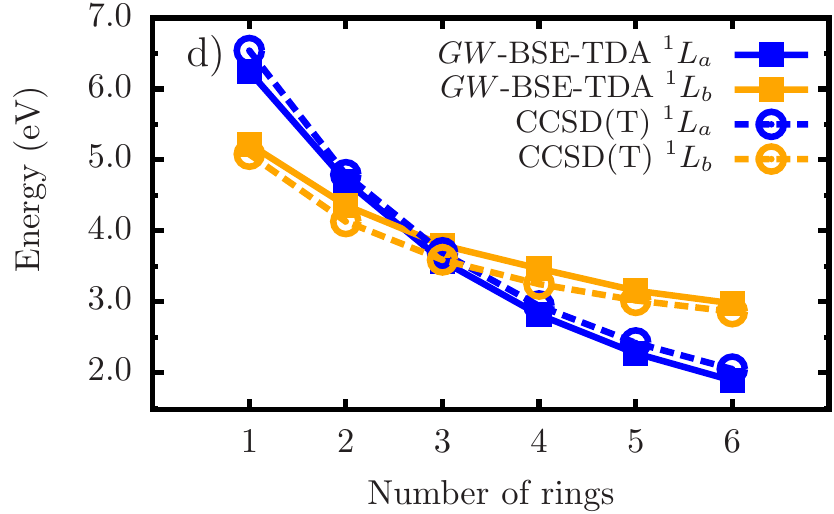}
\end{tabular}
\caption{Low-lying singlet excitations of acenes calculated with TD-OTRSH-PBE and $G_0W_0$-BSE@OTRSH-PBE in panels a and b;
\sla\ and \slb\ excitation energies, with blue and orange lines respectively, are compared to CCSD(T) references from Refs.~\cite{lopata_excited-state_2011} and \cite{silva-junior_benchmarks_2010} (dashed lines).
The corresponding excitations with the TDA at the TDDFT and $GW$-BSE theories are shown in panels c and d.
}
\label{fig:la-lb-vs-ccsdt}
\end{figure*}

In Figure~\ref{fig:mae-exct-tda}a, we show the mean signed deviation~($\text{MSD}= 1/N_i \sum_i^{N_i} E_i - E_i^{\mathrm{ref}}$), relative to CCSD(T)~\cite{lopata_excited-state_2011,hajgato_benchmark_2009}, of representative TDDFT-RSHs explored in this study.
Our TD-OTRSH-BNL results are in excellent agreement with those reported in Refs.~\cite{kuritz_charge-transfer-like_2011}~and~\cite{lopata_excited-state_2011}.
In addition, we find that TD-OTRSH-PBE low-lying singlets are within $0.05$~eV of the corresponding TD-OTRSH-BNL excitations.
In fact, as previously discussed~\cite{lopata_excited-state_2011,moore_charge-transfer_2015}, the performance of TD-OTRSH based on BNL or PBE for \sla\ and \slb\ relative to CCSD(T) is very consistent: for both approaches, \sla\ is within 0.1--0.2~eV of the reference, but \slb\ presents larger discrepancies ($\sim 0.4$ eV), as does the \slb$ - $\sla\ gap, which is within $0.6$~eV. 

\subsection{Predicting the low-lying excitations of the acenes with $GW$-BSE}
\label{gw-bse}
Having reviewed the accuracy of TDDFT-RSH for the \sla\ and \slb\ excitations relative to CCSD(T), we now discuss the $GW$-BSE results, focusing on the sensitivity to the underlying $GW$ starting point.
In Figure~\ref{fig:mae-exct-tda}b, we show the calculated MSD, as defined in the previous section, of representative $GW$-BSE approaches studied here.
Consistent with previously-reported $GW$ results on the charged excitations of the acenes (see for instance Refs.~\cite{bruneval_benchmarking_2013}, \cite{gallandi_accurate_2016} and \cite{rangel_evaluating_2016})  hybrid starting points for $G_0W_0$ or self-consistent $GW$ approaches are required to predict accurate excitations within $GW$-BSE; relatively low MSDs are found within $G_0W_0$-BSE@BHLYP and ev$GW$-BSE, in agreement with recent works~\cite{bruneval_systematic_2015,jacquemin_benchmarking_2015,jacquemin_00_2015}.
As hypothesized in Ref.~\cite{refaely-abramson_quasiparticle_2012}, the OTRSH starting point is superior.
In particular, we highlight that while the OTRSH-PBE starting point yields neutral excitation energies with accuracies similar to other starting-points for aromatic hydrocarbons, OTRSH-PBE leads to markedly improved triplet energetics for the molecules studied here, as discussed later. 
As shown in Figure~\ref{fig:mae-exct-tda}b, for the $GW$-BSE schemes considered here, the \slb\ state is predicted within 0.1--0.2~eV, whereas \sla\ is underestimated by at least 0.4~eV; additionally, the \slb$ - $\sla\ gap is underestimated by $\sim 0.6$~eV independent of $GW$-BSE scheme.
The rather poor performance of both $GW$-BSE and TDDFT approaches, in the context of neutral low-lying singlet and triplet excitations of acene molecules, can be remedied by the Tamm-Dancoff approximation, as discussed next.

\subsection{The role of the Tamm-Dancoff Approximation within TDDFT and $GW$-BSE}
\label{tda}

\begin{table}[h]
\begin{tabular}{lccccccc}
\hline\\
\multicolumn{6}{l}{ $\boldsymbol{L_a}$ \bf states}\\
& \multicolumn{3}{c}{Singlets} & \hspace{0.05 in} & \multicolumn{3}{c}{Triplets}\\\cmidrule{2-4}\cmidrule{6-8}
& BSE & TDA & CCSD(T)$^*$ && BSE & TDA & CCSD(T)$^*$\\
Benz.
& 5.89 & 6.13 & 6.54$^b$ 
&& 3.60 & 3.94 & $4.26 \pm 0.11^{b,c}$ \\
Naph.
& 4.34 & 4.59 &  $ 4.81 \pm 0.02^{a,b}$
&& 2.65 & 2.93 & $3.20 \pm 0.11^{b,c}$\\
Anth.
& 3.38 & 3.63 & $3.68 \pm 0.02^{a,d}$
&& 2.01 & 2.29 & $2.41 \pm 0.07^{c,d}$\\
Tetra.
& 2.42 & 2.72 & 2.94$^a$ 
&& 1.08 & 1.36 & 1.76$^c$\\
Penta.
& 1.88 & 2.21 & 2.42$^a$ 
&& 0.57 & 0.91 & 1.37$^c$\\
Hexa.
& 1.48 & 1.84 & 2.05$^a$
&& $<0 ^\dagger$ & 0.58 & 1.00$^c$\\
Azu.
& 2.00 & 2.12 & 1.94$^d$ 
&& 1.35 & 1.42 & 2.18$^d$\\
BP 
& 3.65 & 3.82 & 4.09$^d$ && 1.95 & 2.41 & 2.82$^d$\\
DBAn
& 3.48 & 3.69 & 3.91$^d$ &&
1.90 & 2.30 & 2.73$^d$\\
\\
MSD & -0.43 & -0.18  &&& -0.74 & -0.39&\\
MAD &  0.44 &  0.22  &&&  0.74 &  0.39&\\
\hline\\
\multicolumn{7}{l}{$\boldsymbol{L_b}$ \bf states}\\
&\multicolumn{3}{c}{Singlets} & \hspace{0.1 in} & \multicolumn{3}{c}{Triplets}\\\cmidrule{2-4}\cmidrule{6-8}
& BSE & TDA & CCSD(T)$^*$ && BSE & TDA & CCSD(T)$^*$\\
Benz.
& 5.10 & 5.15 & 5.08$^b$ 
&& 4.39 & 4.42 & 4.86$^b$\\
Naph.
& 4.30 & 4.32 & $ 4.19 \pm 0.06^{a,b}$ 
&& 3.76 & 3.79 & 4.09$^b$\\
Anth.
& 3.82 & 3.79 &  $3.58 \pm  0.01^{a,d}$
&& 3.42 & 3.43 & 3.52$^d$\\
Tetra.
& 3.33 & 3.37 & 3.25$^a$ 
&& 3.11 & 3.18 &\\
Penta.
& 3.10 & 3.13 & 3.02$^a$ 
&& 2.79  & 2.83 & \\
Hexa.
& 2.91 & 2.98 & 2.86$^a$
&& $^\dagger$ & 2.66 \\ 
Azu.
& 3.34 & 3.49 & 3.64$^d$ 
&& 2.09 & 2.19 & 2.20$^d$\\
BP 
& 3.57 & 3.60 & 3.50$^d$ && 
2.96 & 3.11 & 3.34$^d$\\
DBAn
& 3.58 & 3.61 & 3.57$^d$ && 
  3.34 & 3.43 & 3.35$^d$\\
\\
MSD    &  0.04 &  0.08  &&& -0.23 & -0.17&\\
MAD    &  0.11 &  0.12  &&&  0.23 &  0.20\\
\end{tabular}\\
\hspace{0.1in}\\
$^*$ CCSD(T) data from the literature: 
$^a$ Ref.~\citenum{lopata_excited-state_2011}, 
$^b$ Ref.~\citenum{silva-junior_benchmarks_2010}, 
$^c$ Ref.~\citenum{hajgato_benchmark_2009} 
, and $^d$ Ref.~\citenum{moore_charge-transfer_2015}.\\
$^\dagger$ the BSE Hamiltonian contains negative eigenvalues (read text).
\caption{Singlet and triplet energetics of representative aromatic hydrocarbons calculated with $GW$-BSE@BHLYP with the full-BSE~(denoted simply BSE above) and the TDA.
We consider benzene (Benz), naphthalene (Naph.), anthracene (Anth.), tetracene, (Tetra.), pentacene (Penta.), hexacene (Hexa.), azulene (Azu.), benzo[$e$]pyrene (BP) and dibenzo[$a,c$]anthracene (DBAn). 
MSD and MAD with respect to CCSD(T) are also shown (see text). All energies are in units of eV.
}
\label{table:hydrocarbons}
\end{table}

The fact that the TDA can improve the description of low-lying neutral excitations of the acenes has been discussed thoroughly in the TDDFT community~\cite{sears_communication:_2011,wang_improving_2008,hirata_time-dependent_1999,peach_overcoming_2012,peach_influence_2011},
and here we find that similar arguments apply to $GW$-BSE.
In Figure~\ref{fig:mae-exct-tda}, we also show the MSD of the calculated low-lying excited states using TDDFT and $GW$-BSE within the TDA with respect to the CCSD(T) reference.
The calculated \sla\ and \slb\ energies of acene molecules within representative $GW$-BSE and TDDFT schemes are shown in Figure~\ref{fig:la-lb-vs-ccsdt}.

{\it Link between TDDFT and triplet instability.}
Within Hartree-Fock~\cite{seeger_selfconsistent_1977} and within DFT~\cite{bauernschmitt_stability_1996}, the stability of the spin-restricted solution against that of the more flexible spin-unrestricted solution requires the positive-definiteness of two matrices (one for singlet final states and one for triplet final states) that are precisely the sum of the blocks $A$ and $B$ [See Eq.~(\ref{eqn:hamiltonian})] used in time-dependent Hartree-Fock (TDHF) or in TDDFT. 
In other words, if either one of the $A+B$ matrices has a negative eigenvalue, then the ground-state singlet solution is unstable against a spin-unrestricted triplet solution.
This is the so-called triplet instability.
Consequently, an unstable or near-unstable spin-restricted ground state implies negative or very small eigenvalues of $A+B$, which in turn produce non-physical or too-small neutral excitations in TDHF or in TDDFT~\cite{casida_charge-transfer_2000,sears_communication:_2011}.
This is why prior work often resorts to the TDA to circumvent the spin-restricted instability situation~\cite{casida_charge-transfer_2000,wang_improving_2008,peach_influence_2011,sears_communication:_2011,peach_overcoming_2012,casida_progress_2012}.
The TDA is thus a practical way to prevent the electronic system from sampling the triplet ground state, which is spuriously too low in energy. 

{\it Link between the BSE and triplet instability.}
In the $GW$-BSE framework, the connection between triplet instability and the BSE matrix (Eq.~\ref{eqn:hamiltonian}) is precisely analogous.
However, the connection cannot be demonstrated as rigorously as for TDDFT.
The BSE, evaluated in the standard fashion~\cite{onida_electronic_2002}, is indeed a combination of  (1) eigenvalues obtained from the dynamical $GW$ self-energy and (2) a kernel, which is an approximate functional derivative of the static $GW$ self-energy, namely the static screened exchange approximation~(SEX)~\cite{hedin_new_1965}.
Additionally, the functional derivative $\delta W / \delta G$ is always neglected in the BSE kernel~\cite{onida_electronic_2002}.
Thus, following the same logic for $GW$-BSE as for TDDFT above, the BSE blocks $A+B$ would then lead to stability problems (if present) in the static screened exchange spin-restricted solution.
If one admits that the $GW$ quasiparticle energies are not far from the static screened exchange energies, the connection between triplet instability and BSE can be understood, but again, not proven.
In situations where the triplet instability occurs or nearly occurs in static screened exchange, the TDA to the BSE may be a good route to obtain meaningful neutral excitation energies.
However, this calls for a direct numerical comparison, which we carry out below. 

{\it Performance of the TDA within TDDFT.} As demonstrated in prior work literature~\cite{wang_improving_2008,peach_overcoming_2012}, the TDA improves the description of the \sla\ singlet and the first triplet \tla, states because these share a similar origin; both are covalent in the valence bond sense, and involve mainly HOMO to LUMO transitions, whereas $L_b$ energies are virtually unmodified with the TDA.
Hence, within the RSH time-dependent approaches used here, the large discrepancy (0.4~eV) between the calculated \slb\ state and CCSD(T) is not improved by the TDA (see Figure~\ref{fig:mae-exct-tda}).
On the other hand, we find that the TDA leads to an improvement in the \sla\ excitation energies in the asymptotic limit of longer acene molecules (see panels a and c of Figure~\ref{fig:la-lb-vs-ccsdt}).
For example, for pentacene TD-OTRSH-PBE predicts \sla~$= 2.18$~eV with TDDFT and $2.48$~eV with the TDA, in outstanding agreement with the CCSD(T) reference value of $2.42$~eV. 
In brief, and in agreement with the literature~\cite{lopata_excited-state_2011,moore_charge-transfer_2015}, TDDFT-TDA with RSH functionals yields highly accurate CT-like \sla\ energetics, but tends to overestimate \slb\ transition energies.  

{\it Performance of the TDA within $GW$-BSE.}
Having reviewed the ability of the TDA within TDDFT to predict the low-lying excitations of the acenes, we now discuss the accuracy of the TDA within $GW$-BSE for these transitions.
Here we expand our discussion to a larger set of aromatic hydrocarbons, including azulene, BP and DBAn (see Figure~\ref{fig:molecules}) which have well-characterized $L_a$ and $L_b$ states~\cite{moore_charge-transfer_2015}.
In Table~\ref{table:hydrocarbons}, we show the calculated singlet and triplet excitations with $G_0W_0$-BSE@BHLYP (with and without the TDA) and mean deviations with respect to CCSD(T), as previously defined.

Similar to TDDFT, in $GW$-BSE we find that, independent of $GW$ self-energy scheme, the \sla\ and \tla\ states are improved within the TDA (by at least $0.2$~eV, see Figure~\ref{fig:mae-exct-tda} and Table~\ref{table:hydrocarbons}).
While triplet energies remain underestimated by $\sim 0.3-0.4$~eV, singlet energies are accurately predicted with $GW$-BSE-TDA, with remaining discrepancies lower than $0.2$~eV.
The TDA also corrects the \sla\ -- \slb\ energy ordering; as shown in Figure~\ref{fig:la-lb-vs-ccsdt} these two states cross at naphthalene when following increasing/decreasing ring number $n$ within the full-BSE (panel b), while the crossing is at anthracene within the TDA (panel d), in agreement to CCSD(T).
In summary, $GW$-BSE within the TDA can predict -- with excellent quantitative accuracy, an MSD better than 0.2~eV -- both ionic CT-like and covalent singlet excitations (such as \sla\ and \slb, respectively) of the acenes and other aromatic-hydrocarbons.

\begin{figure}[h]
\includegraphics{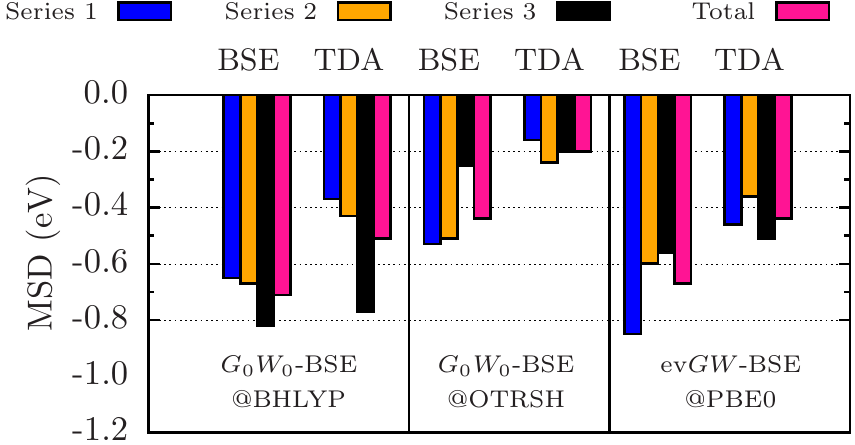}
\caption{
First-triplet excitation energies of organic molecules in Thiel's set (see Figure~\ref{fig:molecules}) calculated with $GW$-BSE are benchmarked against reference data~\cite{silva-junior_benchmarks_2010}.  
The MSD (read text) corresponding to molecules in Series 1 is shown in blue bars, Series 2 in orange bars, Series 3 in black bars and the total in pink bars.
We consider several $GW$-BSE schemes with the full-BSE and the TDA.
}
\label{fig:mae-t1}
\end{figure}

{\it Thiel's set.} We further analyze the accuracy of $GW$-BSE and the TDA for a larger set of triplet excitations. 
We show in Figure~\ref{fig:mae-t1} the MSDs (previously defined) of the calculated first-triplet energies of 20 organic molecules of Thiel's set (all energies are tabulated in the SI).
Again, we consider several representative $GW$-BSE approaches; using the BHLYP and OTRSH-PBE starting points for $G_0W_0$ and ev$GW$. Additionally, we show the MSDs for the molecule categories described in Figure~\ref{fig:molecules}.

With this larger set of excitations, it becomes clear from our calculations that $G_0W_0$@BHLYP and ev$GW$@PBE0, known to perform reasonably well for singlet excitations~\cite{jacquemin_benchmarking_2015,bruneval_systematic_2015}, can present severe errors for triplets, with MSDs of $\sim -0.6$ to $-0.8$~eV, as noticed first in Ref.~\cite{bruneval_systematic_2015} for the BHLYP starting point.
The OTRSH starting point for $G_0W_0$-BSE has a relatively lower MSD of $\sim -0.4$ to $-0.6$~eV, presumably due to the RSH optimal starting point for the underlying $GW$ electronic structure~\cite{gallandi_accurate_2016,rangel_evaluating_2016}, which will be discussed in detail in a separate publication~\cite{samia}.
For all $GW$-BSE approaches studied here, the TDA improves the first-triplet energy, a fact that we discuss in greater depth below.
Further, we note that in agreement with recent work~\cite{jacquemin_benchmark_2017}, $GW$-BSE-TDA approaches predict inaccurate triplet energies (with MSD of $-0.4-0.5$~eV) when using a global-hybrid starting point. 
Importantly, within the OTRSH starting-point, $GW$-BSE-TDA can result in relatively accurate first-triplet energies with a MSD of $-0.19$~eV. 

\begin{figure}
\includegraphics{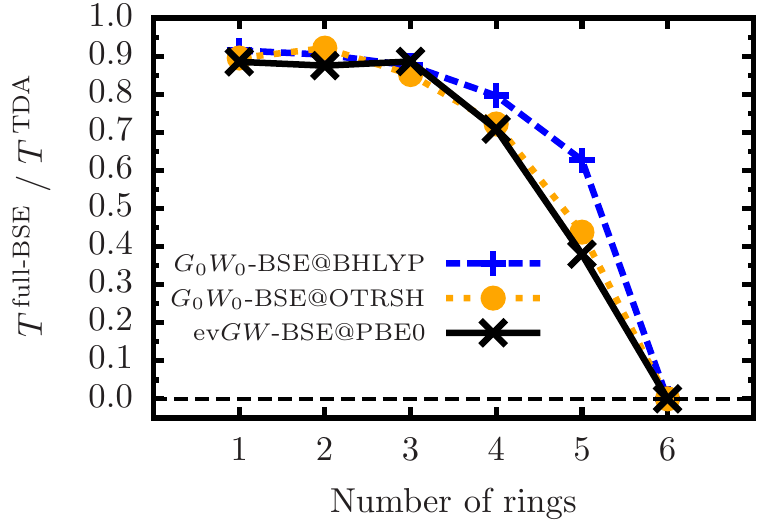}\\
\caption{Ratio of the first triplet energy~($T$) calculated within $GW$-BSE diagonalizing the full BSE Hamiltonian and using the TDA.
Several representative $GW$-BSE schemes are shown: $G_0W_0$-BSE@BHLYP in dashed-blue lines and crosses, $G_0W_0$-BSE@OTRSH-PBE in dotted-orange lines and circles and ev$GW$-BSE@PBE0 in black lines and crosses.
$GW$-BSE predicts a negative triplet energy (shown at zero) for hexacene for all $GW$ schemes used in this work.}
\label{fig:ratio-triplet}
\end{figure}

In order to better understand the superior performance of the TDA within $GW$-BSE for low-lying triplet energies, we show in Figure~\ref{fig:ratio-triplet} the ratio of the triplet energy calculated within the full $GW$-BSE and within the TDA ($T/T^{\mbox{\scriptsize TDA}}$).
When the ratio approaches zero or becomes negative (TDDFT predicts a negative or zero triplet energy), the triplet and its corresponding ground state become unstable, as explained before; hence this ratio acts as a measure of instability~\cite{sears_communication:_2011}.
In this work, we find that the full-BSE and the TDA predict similar triplet energies for benzene and naphthalene (ratio close to 0.9), but the triplet ratio drops to less than zero at hexacene independent of the $GW$ approximation; note that for the PBE starting point to $GW$-BSE, the ratio becomes negative at pentacene (not shown). 
This implies that $GW$-BSE, in disagreement with CCSD(T)~\cite{hajgato_benchmark_2009}, predicts triplet ground states for acenes larger than pentacene.
In analogy to TDDFT, this may be a result of instabilities in the corresponding $GW$-BSE triplet and ground states; we leave the evaluation of stability conditions in the $GW$-BSE states to future work.
 
As mentioned above, triplet instabilities are well-known and documented in Hartree Fock and TDDFT theories~\cite{sears_communication:_2011,peach_overcoming_2012,peach_influence_2011,hirata_time-dependent_1999,casida_progress_2012},
and $GW$-BSE is similarly affected, as shown here and in Ref.~\cite{zimmermann_influence_1970}.
In TDDFT---as in configuration interaction singles (CIS) theory, which mixes only single Slater determinants and is the minimal post-Hartree-Fock method capable of predicting physical excited states---triplet instabilities are overcome with the TDA~\cite{casida_progress_2012}. 
Not surprisingly, in $GW$-BSE the TDA also overcomes triplet instabilities, as we document here, in a manner analogous to TDDFT for molecules.

Finally, we briefly comment on the performance of the BSE among other {\it two-particle Green's function methods}.
In this context, the algebraic diagrammatic construction~[ADC(2)]~\cite{trofimov_efficient_1995} and the second-order polarization propagator~(SOPPA)~\cite{nielsen_transition_1980} methods are efficient Green's function methods which give access to the neutral excitations of molecules.
The accuracy of representative variants of ADC(2) has been studied for the low-lying singlet excitations of the acenes, from naphthalene to hexacene, in Ref.~\cite{knippenberg_low-lying_2010}.
Comparing these results to the CCSD(T) reference used in this work, we find an MSD for the ADC(2) variants of $-0.46$ and $0.36$~eV (or lower) for \sla\ and \slb\ respectively, relatively higher than that of the best $GW$-BSE method used in this work ($GW$-BSE-TDA@OTRSH with an MSD of $-0.13$ and $0.17$~eV for the same excitations)~\footnote{
The calculated singlet energies of the acenes with strict ADC(2) [ADC(s)-s] in Ref.~\cite{knippenberg_low-lying_2010} are:
4.57, 4.00, 3.61, 3.36 and 3.18~eV for \slb, from naphthalene to hexacene respectively, and
5.09, 3.87, 3.04, 2.46 and 2.05~eV for \sla.
The extended variant [ADC(2)-x] yields poorer results for these excitations~\cite{knippenberg_low-lying_2010}.
Similar values for the singlet excitatios of naphthalene are found in Ref.~\cite{helmich_pair_2014} with a smaller TZVP basis. 
We use these results to calculate the MSD of ADC(2) with respect to the CCSD(T) references (shown in the SI).}
. 
In Refs.~\cite{packer_new_1996,sauer_performance_2015}, the lowest singlet excitations of small acenes have been calculated with SOPPA-based methods, among which the original SOPPA and SOPPA(CCSD) perform best.
The MSD with respect to CCSD(T) is 0.3--0.8~eV for \sla\ and \slb\ and 0.4--0.5~eV for the first triplet energies of benzene and naphthalene~\footnote{
We have compiled the calculated energies with SOPPA-based methods from Ref.~\cite{sauer_performance_2015} with the aug-cc-TZVP basis, which are also found in Ref.~\cite{packer_new_1996} with a smaller basis set.
The \sla\ energies of benzene and naphthalene calculated with SOPPA are 5.91 and 4.19, respectively, and 5.77 and 3.97 with SOPPA(CCSD). 
The corresponding \slb\ energies are 4.63 and 3.78~eV with SOPPA and 4.43 and 3.49, with SOPPA(CCSD).
The first triplet energies within SOPPA and the cc-pVTZ basis are 3.73 and 2.68~eV for benzene and naphthalene, respectively;
the corresponding values within SOPPA(CCSD) are 3.56 and 2.44~eV
}
.
Moreover, for the triplet excitations in the Thiel's set, SOPPA and SOPPA(CCSD) yield MSDs with respect to the BTE of $-0.45$ and $-0.54$~eV, respectively~\cite{sauer_performance_2015},
which again is relatively high compared to $G_0W_0$-BSE-TDA@OTRSH in this work (with an MSD of only $-0.2$~eV).
The superior performance of $GW$-BSE-TDA for these sets of excitations therefore situate the BSE as an efficient and accurate alternative to many traditional approximate methods in quantum chemistry.

\section{Conclusions}
\label{conclusions}

In summary, we have benchmarked $GW$-BSE with CCSD(T) for neutral excitations of aromatic hydrocarbons and heterocycles, including the challenging $L_a$ and $L_b$ excitations heavily documented in prior work with TDDFT.
We first explored the accuracy of approximations to $GW$-BSE and found that $G_0W_0$-BSE@OTRSH can yield accurate triplet and singlet excitations, sometimes outperforming other highly-accurate approaches such as ev$GW$-BSE and $G_0W_0$-BSE@BHLYP.
In particular, for aromatic hydrocarbons, the above mentioned $GW$-BSE methods can predict accurate \slb\ energetics but generally present significant errors for the \sla\ states.
This problem is remedied by using the TDA, which leads, as it does with TDDFT, to a better overall performance, overcoming triplet instabilities, improving triplet energetics, and capturing quantitatively both the charge-transfer-like $L_a$ and covalent $L_b$ singlet excitations of aromatic cyclic compounds.

\section{Supplemental Material}
In the Supplemental Material we tabulate the calculated neutral excitations of the acenes and other organic molecules of the Thiel's set.

\acknowledgments
This work was supported by the Center for Computational
Study of Excited State Phenomena in Energy Materials at
the Lawrence Berkeley National Laboratory, which is funded
by the U.S. Department of Energy, Office of Science, Basic
Energy Sciences, Materials Sciences and Engineering Division
under Contract No. DE-AC02-05CH11231, as part of the
Computational Materials Sciences Program. This work is also supported by the Molecular Foundry through the U.S. Department of Energy, Office of Basic Energy Sciences under the same contract number. 
We acknowledge the use of computational resources at the National Energy Research Scientific Computing Center (NERSC).
F. Bruneval acknowledges the Enhanced Eurotalent program and the France Berkeley Fund for supporting his sabbatical leave in UC Berkeley.
\bibliography{paper}
\end{document}